\documentclass[10pt,conference,compsocconf]{IEEEtran}

\usepackage{times}
\usepackage{caption}
\usepackage{pifont}

\usepackage[cmex10]{amsmath}
\usepackage{url}

\usepackage{amssymb}
\usepackage{paralist}

\usepackage{array}	
\usepackage{float}

\usepackage{stfloats}

\usepackage{color} 
\definecolor{grund}{gray}{.95}  

\usepackage{graphicx}

\usepackage{caption}
\begin{document}

%\title{Identifying Long-Term Risks of the Internet of Things}

\title{\textbf{\Large Long-Term Risks of IoT Devices: The Case of the Smart Fridge\\[0.2ex]}}

\author{
\IEEEauthorblockN{~\\[-0.4ex]\large Erik Buchmann$^{a,b}$\\[0.3ex]\normalsize}
\IEEEauthorblockA{
$^{a}$ Dept. of Computer Science, Leipzig University, Germany\\
$^{b}$ Center for Scalable Data Analytics and Artificial Intelligence Dresden/Leipzig, Germany\\
Email: {\tt buchmann@informatik.uni-leipzig.de}}
}

\maketitle

%%%%%%%%%%%%%%%%%%%%%%%%%%%%%%%%%%%%%%%%%%%%%%%%%%%%%%%%%%%%%%%%%%%%%%%%%%%%%%

\begin{abstract}

Replacing conventional devices with smart ones has many advantages, e.g., a seamless integration of physical objects into the user's digital environment or improved modes of use. 
However, if a conventional device is replaced by a smart device, its IT components can cause risks, that shorten the life of the device. 
Such risks stem from different life cycles of embedded soft- and hardware, libraries and protocols used, and the IT ecosystem required.
This is problematic, because many conventional household appliances, say, a fridge or TV, have a much longer life span than typical IT equipment. 
In this paper, we use a systematic approach to identify long-term risks for the operational life span of a smart fridge. 
In particular, we identify 8 different use cases of three typical smart fridges, e.g., cooling or managing "best before" dates. 
We model the IT ecosystem needed to run these use cases, and we inspect each asset in this ecosystem for potential long-term risks.
We found that even cooling, the most basic use case, is at risk in the long run. This is because the setting cooling parameters may depend on parts of the IT ecosystem that are not under the user's control.  
On the other hand, we did not find any risk that may lead to harm of the category "threatening". Our findings on the smart fridge can be generalized to other smart devices easily. 

\end{abstract}

\renewcommand\IEEEkeywordsname{Keywords}
\vspace{0.25cm}

\section{Introduction}
\label{sec:intro}

Advances in hard- and software have led to the trend to add sensors, computational resources and communication interfaces to traditional consumer products, and to connect them over the Internet to cloud services where an artificial intelligence approach interprets voice commands or enhance user experience. Together, such smart devices form the Internet of Things (IoT)~\cite{atzori2010internet}. In many cases, smart devices stem from non-smart predecessors. For example, a modern smart refrigerator looks and feels 
much like a classic non-smart refrigerator with some extras, e.g., remote control and expiration management for perishable foods.

Smart devices allow consumers to create smart homes with devices that can be controlled remotely via smartphone, adapt to the user's habits, and provide convenient services locally or on the Internet. However, media provide evidence that smart devices might come with operational risks that occur well after the time of purchase. 
With a familiar non-smart device in mind, customers may not expect risks like Examples 1-3, when choosing a smart device.

\textbf{Example 1:} 
%Different software and hardware lifecycles. 
%\textsl{The hardware of a typical fridge has an operational life-span of 14 years~\cite{fridgelife2022}. However, it is unlikely that the software of a smartphone purchased 14 years from now will be compatible with a smart fridge purchased today.}
\textsl{The software of a smart device may have an operational life-span that is much shorter than the life-span of its hardware~\cite{zdankin2020towards}. For example, without regular functional and security updates, a smart TV soon becomes useless~\cite{androidtv}.}

\textbf{Example 2:} 
%Loss of control. 
\textsl{Smart devices may be tied to a cloud service. For example, after a third-party service provider ceased its business, tens of thousands smart Internet radios became non-functional~\cite{smartradio} without warning in advance.} 

\textbf{Example 3:} 
%Compliance or legislation. 
\textsl{Changes in the legislation may prohibit the use of smart devices after years of operation. In Germany, for example, a child's smart toy~\cite{cayla2022} has been forbidden as a spying tool, three years after it had been introduced to the market, because it was not visible that the toy sends voice recordings into the cloud. 
}

In order to make smart devices accessible for risk management, a comprehensive catalog of potential risks is required. 
It is challenging to find a research method that delivers such a catalog. For example, the results of a study~\cite{tanczer2018emerging} depend on the insights of the study participants. 

In this paper, we use smart fridges as a use case to compile a comprehensive set of long-term risks that are (a) specific for the smart fridge, i.e., do not exist for conventional fridges, and (b) may materialize years after the fridge has been purchased. 
We define our problem statement as follows: 

\textit{Which specific risks for the continued long-term use of a smart fridge may appear after purchase, but cannot be expected from a conventional fridge?}

We call a fridge a ``smart fridge'', if it contains computational capabilities and data links, which are not essential for the primary function "cooling food products". 
By ``long-term'', we refer to an operational life of $>$10 years, which can be expected from a fridge's hardware~\cite{fridgelife2022}.
Intuitively, this may be the expectation of a customer replacing a broken fridge. 

In this paper, we adapt our research method from~\cite{buchmann2020identifying} to methodically derive such long-term risks for a smart fridge in a domestic environment. 
We have identified \textsl{compliance risks} resulting from changing local, national or international rules, \textsl{economic risks} from future business decisions of the organizations involved, and \textsl{operational risks} considering the technical perspective of operating a smart fridge together with its IT ecosystem for more than 10 years. Due to our methodical approach, we consider our set of risks to be complete for this application scenario. We think that it can be easily adapted to similar scenarios.

The paper is structured as follows: In Section~\ref{sec:related}, we briefly review related work. In Section~\ref{sec:method}, we sketch our approach to identify long-term risks of smart fridges. In Section~\ref{sec:results}, we use this approach to obtain our set of risks. Section~\ref{sec:conclusion} concludes.

\section{Related Work}
\label{sec:related}

\begin{figure*}[htb]
	\centering
	\includegraphics[width=1.3\columnwidth,trim=0.5cm 1.2cm 0.0cm 1.9cm, clip]{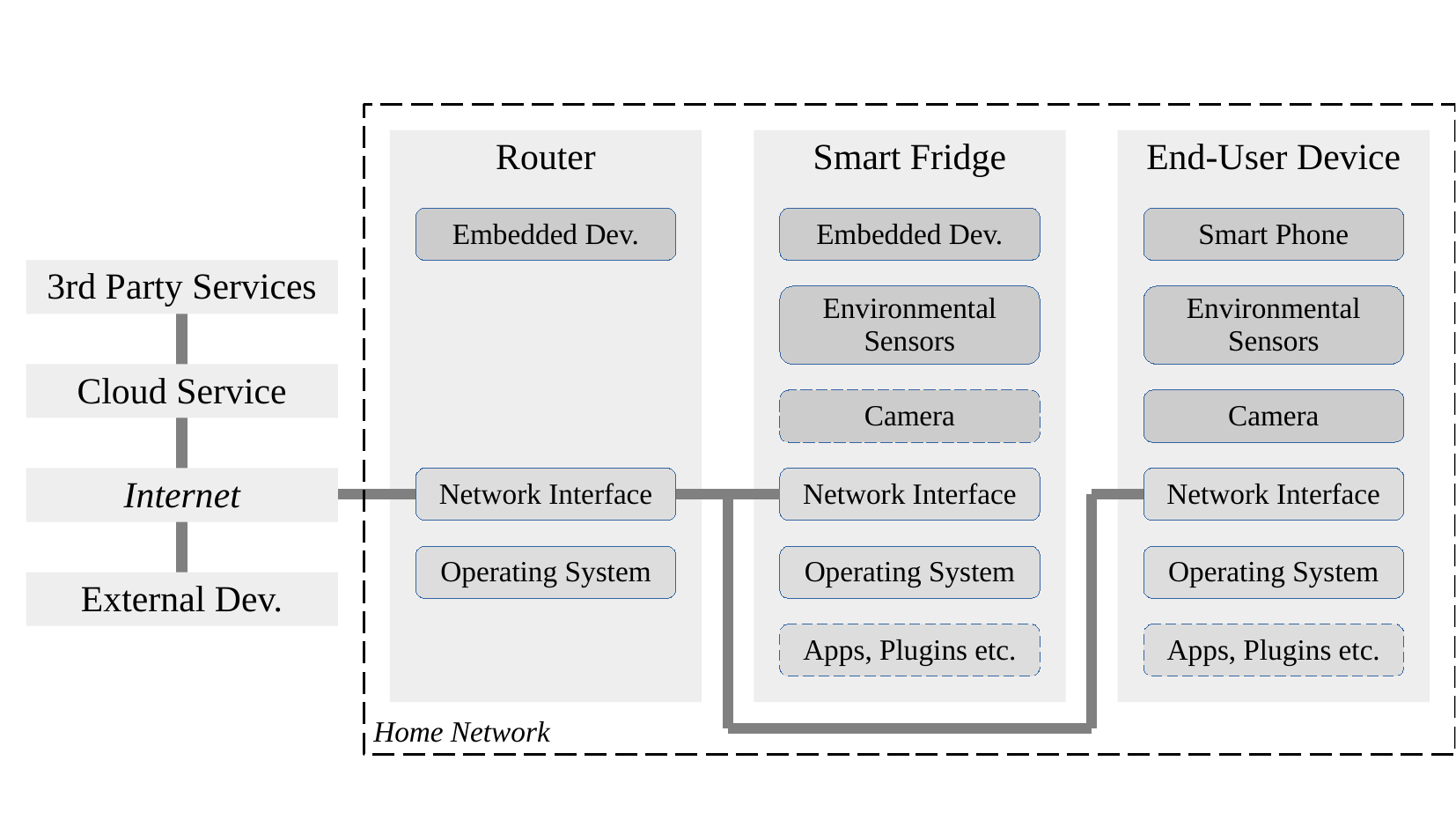}
	\caption{IT Architecture}\label{img:archi}
\end{figure*}

This section summarizes methods, standards and findings related to our work.
%
% \subsection{Design Science Research}
\textbf{Design science research}~\cite{hevner2010design} is a method to design an artefact from a knowledge base, and evaluate and improve it in several rounds. Each round is divided into three cycles:
The \textit{relevance cycle} specifies and refines the use cases needed to construct the artefact and evaluate its applicability. 
The \textit{rigor cycle} builds a knowledge base from literature and experience that is needed to evaluate the novelty and the research contribution of the artefact. 
The central \textit{design cycle} iterates between building and evaluating the artefact, based on information from the other cycles. 

% \subsection{BSI-Standard 200-3} 
The \textbf{BSI-Standard 200-3}~\cite{bsi200-3en} is based on IT-Grundschutz. It defines a process that allows organizations to assess their information security risks. In particular, the standard describes how to identify, classify, consolidate, assess and treat risks. Our concern is risk identification. In this respect, the standard distinguishes risks that arise from \textsl{elementary threats}, e.g., fire, theft, misconfiguration or manipulation, and \textsl{specific threats} from specific use cases. The risk identification starts with the modelling of the use cases. 
The risk catalog is then compiled from the consolidated risks of the individual IT assets, which are part of the model.

% \subsection{Risk Analyses for IoT}
Advances in technology call for \textbf{risk analyses} before adoption. However, risk analyses typically use a descriptive research approach, focus on the current situation and/or have a narrow perspective, e.g., on current IT security or return on investment. For example, \cite{alladi2020consumer} reviews vulnerabilities of smart devices in the consumer market. The risk assessment approach in \cite{aydos2019assessing} considers the management of risk over the past two years, but does not project into the future, e.g., when security breaches remain untreated for a discontinued product. A study~\cite{tanczer2018emerging} provides a holistic view on future IoT risks, 
but a standardized questionnaire cannot provide a complete overview on future risks. In consequence, existing approaches that deal with IoT risks during the operational life of the device~\cite{garcia2015comprehensive} \cite{hernandez2016army} do not consider that vendors may loose interest in supporting discontinued products, or that it will be hard to find experts to maintain outdated technology.
\cite{zdankin2020towards} uses threat models to assess risks due to discontinued services, breaking updates, trade conflicts, etc., but it remains unclear if this risk assessment is exhaustive for the devices in question. 
In~\cite{buchmann2020identifying}, we have defined a research method to identify long-term risks that are specific for smart devices. Because this method is fundamental for our paper, we will explain it in more detail in the next section. 

% \subsection{Long-Term Digital Preservation}
The \textbf{long-term preservation of digital goods} has been extensively discussed~\cite{digital2015digital} in the past years. The risks for digital content~\cite{vermaaten2012identifying} overlap with the risks of using an outdated smart device in a modern environment.
Examples are media obsolescence and format obsolescence~\cite{vermaaten2012identifying}, i.e., the digital object cannot be read with current devices due new media or new formats. Security properties have been established with protocols that are insecure now~\cite{gladney2004trustworthy}. Digital objects, such as dynamic web pages~\cite{truman2016web} or computer games~\cite{andersen2011games}, require a complex execution environment.

\section{How to Identify Long-Term IoT Risks}
\label{sec:method}

In this section, we briefly describe our research method, which we have developed in~\cite{buchmann2020identifying}.
Our method adapts BSI-Standard 200-3~\cite{bsi200-3en} so that it creates the knowledge base and designs a risk catalog that fits into relevance and design cycle of Design Science Research~\cite{hevner2010design}. We use research literature to foster the rigor cycle. 

For this paper, we have extended two aspects of \cite{buchmann2020identifying}: We explicitly refer to an operational scenario (in our case: a domestic environment) to assess the extent of potential harms and damages. Furthermore, we do not add assets to our infrastructure model that do not add specific risks for the smart device in question, e.g., the Internet router or the electricity supplier.
In particular, we use the following steps:
\begin{compactenum}
\item Select typical devices and identify the use cases for these devices in a given scenario. Model its IT infrastructure. 
\item Determine under which conditions each asset in this infrastructure operates as intended by the use cases.
\item Append this condition to the risk set, if it is not present at purchase and doesn't exist for non-smart devices.
\item Assess the harm the risks could cause, and consolidate risks that are identical for multiple assets. 
\item Back up each individual risk by literature. 
\end{compactenum}

% For illustration, we apply this approach to Example~3 from the introduction. \textit{Step 1:} The generic IT infrastructure for the smart security camera contains, among other things, a data connection between the smart device and a cloud service provider in the UK. This is because the security camera vendor has outsourced the burglar alert into the cloud. The connection transports personal data, e.g., videos of humans. \textit{Step 2:} One required condition is that the data transfer is legal - depending on the legislation. \textit{Step 3:} A common non-smart security camera uses a local storage system, not needing a legal authorisation for cross-border data transfers. \textit{Step 4:} ``Changing privacy legislation for data transfers into other countries'' is not an elementary risk. \textit{Step 5:} A body of literature can be identified, discussing the risks of changes in the privacy regulations for transferring data to a UK cloud, e.g.,~\cite{mccullagh2017brexit}. Thus, we have identified ''changing privacy legislation'' as a plausible risk for \textsl{any} smart device that uses such an IT infrastructure to transfer personal data. 

\section{Long-term Risks of a Smart Fridge}
\label{sec:results}

In this section, we apply our research method from Section~\ref{sec:method} to identify a comprehensive set of long-term risks associated with smart fridges. 
% In this section, we  use our research method to identify a comprehensive set of long-term risks for smart fridges. 

\textbf{Operational scenario:} We base our analysis on a domestic environment, where the fridge stores perishable food that needs cooling and has a limited economic value. 
The user of the fridge values the user experience more than privacy and likes to use all technical possibilities of the digital services offered by the smart fridge. 
The user can be expected to detect spoiled food, but does not possess the IT-security knowledge needed to detect cyber attacks on the smart fridge. 
Figure~\ref{img:archi} illustrates the IT architecture for this scenario.

% % % % % % % % % % % % % % % % % % % % % % % % % % % % % % % % % % % % % % % % % % % % % % % % % % % % % % % % % % % % % % % % % % % % % % % % 
\subsection{Device Selection and Use Cases}
According to Step~1 of our research method, we select three typical devices from the category "Smart Fridge":

\begin{enumerate}
\item Bosch KGN36HI32
\item Samsung RF27T5501SG
\item LG GSX960NEAZ
\end{enumerate}

The Bosch KGN36HI32 can be controlled via the Bosch Home Connect platform, which connects to Amazon Alexa and other voice assistants and sends temperature alarms to the user's smartphone. It is equipped with internal cameras, that monitor the cooled food products. 
The Samsung RF27T5501SG provides similar technical features as the Bosch fridge, but uses the Samsung product family: It contains a Samsung Family Hub and connects to a voice assistant called Bixby. In addition, it provides a large LCD screen with apps and Internet access via WLAN, and an ice dispenser.
The LG GSX960NEAZ provides the smallest set of smart features: It only controls fridge parameters and functions, such as defrost and alerts, via LG smartphone app. But it does not contain cameras, smart home hubs or LCD screens. 

\begin{table}[htb]
		\caption{\label{tab:usecase} \uppercase{Categories of Use Cases}}
	\centering
		\begin{tabular}[h]{|c|p{1.4cm}|p{5.6cm}|}\hline
		\textsl{Id} & \textsl{Name} & \textsl{Description} \\\hline
U1 & Cooling & Storing and cooling food products.\\
U2 & Monitoring & Monitoring the food storage via camera.\\
U3 & Management & Managing food expiration and shopping lists.\\
U4 & Shopping & Replenish food storage.\\
U5 & Multimedia & Playing music, TV streams, Internet access.\\
U6 & Remote & Remote control of cooling and alarms.\\
U7 & Apps & Other apps, e.g., for searching wine temperatures.\\
U8 & Updates & Functional upgrades or security updates.\\
			\hline		
		\end{tabular}
\end{table}

To obtain typical use cases for smart fridges, we have browsed the manuals and web pages of our selected devices. 
Table~\ref{tab:usecase} lists all use cases we have identified. \textbf{Cooling} (U1) is the traditional use of a fridge. \textbf{Monitoring} (U2), \textbf{management} (U3), \textbf{shopping} (U4) and \textbf{remote control} (U6) refer to typical domestic requirements, which are now enhanced by digital services. Smartphone apps control cooling parameters and various alarms (opening, temperature, humidity), look inside the fridge via cameras, and might also identify food products that are used up or are close to its expiration date. If the smart fridge is part of a larger smart-home solution, it typically serves as a \textbf{multimedia} (U5) hub to deliver audio and video streams to connected devices.
Some smart fridges allow \textbf{further apps} (U7), e.g., to manage recipes, to search for optimal wine temperatures or to browse the Internet. \textbf{Updates} (U8) are important to maintain the security and functionality of the smart fridge during its operational life.

% % % % % % % % % % % % % % % % % % % % % % % % % % % % % % % % % % % % % % % % % % % % % % % % % % % % % % % % % % % % % % % % % % % % % % % % 
\subsection{IT Infrastructure Model}

\begin{table}[H]
		\caption{\label{tab:devices} \uppercase{Categories of Devices}}
	\centering
		\begin{tabular}[h]{|c|p{1.8cm}|p{5.2cm}|}\hline
		\textsl{Id} & \textsl{Name} & \textsl{Description} \\\hline
A1 & Smart Fridge & The smart fridge.\\
A2 & End-User Dev. & Smartphone (WLAN), TV, smart speaker, etc.\\
A3 & Cloud & Digital fridge services on the Internet.\\
A4 & 3rd Party Serv. & Digital smart home services on the Internet.\\
A5 & External Dev. & Smartphone (LTE) or tablet (LTE).\\
		\hline		
		\end{tabular}
\end{table}

Table~\ref{tab:devices} lists all categories of devices or appliances needed to run the use-cases U1-U8. 
The \textbf{smart fridge} (A1) contains an embedded computing device with network interface and operating system. It may or may not also contain further plugins and apps, e.g. a web browser. Some smart fridges are equipped with internal cameras that monitor the stored food products. Any smart fridge we are aware makes use of sensors to monitor parameters, such as temperature and moisture. 
Both \textbf{end-user device} (A2) and \textbf{external device} (A5) are used to control any smart, digital service the fridge offers. Respective devices include laptops, smartphones, tablets, smart TVs or smart speakers. The difference between A2 and A5 is that the external device connects via LTE, i.e., it uses a network connection that leaves the home WLAN. Thus, we need to model it separately.
A3 refers to a \textbf{cloud service} that is bundled with the smart fridge, and provides services tailored to the fridge. For example, Bosch KGN36HI32 connects to Bosch Home Connect. In contrast, A4 means any other \textbf{3rd-party service}, e.g., a smart-home system, a voice assistant or a content-delivery network from a third-party cloud. Since our focus is on the smart fridge, we leave aside the router.

\begin{table}[H]
		\caption{\label{tab:data} \uppercase{Categories of Data}}
	\centering
		\begin{tabular}[h]{|c|p{1.5cm}|p{5.5cm}|}\hline
		\textsl{Id} & \textsl{Name} & \textsl{Description} \\\hline
D1 & Sensor data & Video, audio, temperature, power consumption.\\
D2 & App data & Data from apps installed on the smart fridge.\\
D3 & Metadata & Timestamps, soft- and hardware versions.\\
D4 & Configuration & Parameters, credentials, certificates.\\
D5 & Telemetry & Device behavior, log information.\\
D6 & Op. System & Software libs, updates, operating system data.\\
		\hline		
		\end{tabular}
\end{table}

Use cases U1-U8 require the smart fridge to manage and share 6 categories of data with devices A1-A5, as shown in Table~\ref{tab:data}.
\textbf{Sensor data} (D1) includes any information delivered by internal sensors of the fridge, e.g., video streams from an internal camera or the temperature in the wine compartment. 
\textbf{App data} (D2) refers to data managed by the various kinds of apps executed on the smart fridge. Examples are the user's shopping lists, multimedia data from external parties or expiration dates. 
\textbf{Metadata} (D3) is any information produced by the operation of smart services. Examples include version numbers, timestamps or patch levels of software libraries. 
\textbf{Configuration} (D4) data stores parameters about how the use cases should work. This means cooling parameters as well as WLAN credentials or HTTPS certificates. 
\textbf{Telemetry} (D5) means any information that is typically part of the log file of the smart fridge, e.g., internal errors, defrost times, power outage, and the like.
\textbf{Operating System} (D6) refers to the program code of the operating system and its apps, updates, patches, libraries, etc.

\begin{table}[H]
		\caption{\label{tab:orga} \uppercase{Categories of Organizations}}
	\centering
		\begin{tabular}[h]{|c|p{2.5cm}|p{4.5cm}|}\hline
		\textsl{Id} & \textsl{Name} & \textsl{Description} \\\hline
O1 & User & The user of the smart fridge.\\
O2 & Vendor & The manufacturer of the smart fridge.\\
O3 & Cloud Provider & The operator of the cloud service.\\
O4 & 3rd Party Provider& External cloud service providers.\\
O5 & Other 3rd Parties & Other services.\\
		\hline		
		\end{tabular}
\end{table}

The devices are operated by different parties, as shown in Table~\ref{tab:orga}. 
Since our problem definition focuses on specific risks for a smart fridge, we leave aside the parties that might cause generic risks. Such parties are the Internet provider, the LTE provider or the electricity supplier. 
Large companies, such as LG, have their own cloud infrastructure and cloud services, like voice assistants used by the smart fridge. Thus, O2 and O3 can be the same organization.

\begin{table}[H]
		\caption{\label{tab:network} \uppercase{Categories of Network Connections}}
	\centering
		\begin{tabular}[h]{|c|c|c|c|l|}\hline
		\textsl{Id} & \textsl{Devices} & \textsl{Int.} & \textsl{Pers.} & \textsl{Description} \\\hline
C1 & A1-A2 & \ding{51} & \ding{51} & smart fridge -- end-user device\\
C2 & A2-A3 & \ding{55} & \ding{51} & end-user device -- cloud\\
C3 & A1-A3 & \ding{55} & \ding{51} & smart fridge -- cloud\\
C4 & A3-A4 & \ding{55} & \ding{51} & cloud -- 3rd party cloud\\
C5 & A1-A5 & \ding{55} & \ding{51} & smart fridge -- external device\\
		\hline		
		\end{tabular}
\end{table}

Table~\ref{tab:network} contains all categories of network connections we need to consider. Note that all connections are bi-directional. 
Column "Int." indicates that a connection transfers data within the home WLAN. 
"Pers." means that a connection might transfer data related to the activities or habits of a person. 

\begin{table}[H]
		\caption{\label{tab:matrix} \uppercase{Asset Matrix}}
	\centering
		\begin{tabular}[h]{|c|c|c|c|c|}\hline
		\textsl{U. C.} & \textsl{Data} & \textsl{Devices} & \textsl{Connections} & \textsl{Orga.} \\ \hline
U1 & D1, D3-D5 & A1 &  & O1\\
U2 & D1-D4 & A1, A2, A5 & C1, C5 & O1\\
U3 & D2-D4 & A1-A3, A5 & C1-C3, C5 & O1, O3\\
U4 & D2-D4 & A1, A2, A4, A5 & C1, C4, C5 & O1, O4\\
U5 & D2-D4 & A1-A5 & C1-C5 & O1, O4\\
U6 & D1-D5 & A1-A3, A5 & C1-C3, C5 & O1, O3\\
U7 & D1-D5 & A1-A5 & C1-C5 & O1, O3-O5\\
U8 & D2-D6 & A1, A3  & C3 & O2\\
		\hline		
		\end{tabular}
\end{table}

After having defined the categories of use cases, devices, data, organizations and network connections we need to consider, we can define an asset matrix (cf.~Table~\ref{tab:matrix}). The asset matrix tells which use case is tied to which IT asset.
U1 (Cooling) is the only use case that does not need network connections, other devices or other organizations. All other use cases depend on an operational IT ecosystem.

% % % % % % % % % % % % % % % % % % % % % % % % % % % % % % % % % % % % % % % % % % % % % % % % % % % % % % % % % % % % % % % % % % % % % % % % 
\subsection{Potential Harms and Damages}

Table~\ref{tab:usecase} allows us to devise four categories of potential harm, as shown in Table~\ref{tab:harm}. The categories are in line with~\cite{bsi200-2}. Potentially, a smart fridge may produce threatening physical or financial damages, e.g., from spoiled food or a violation of legal regulations. Negligible harm could be a brief interruption or malfunction of digital or cooling services. 

\begin{table}[H]
		\caption{\label{tab:harm} \uppercase{Categories of Potential Harm}}
	\centering
		\begin{tabular}[h]{|c|p{6.0cm}|}\hline
		\textsl{Category} & \textsl{Examples} \\\hline		
negligible & Food spoils a bit earlier, digital services are unavailable for a short time, isolated false alarms.\\
limited & Fridge contents spoils, services are unavailable for some time, many false alarms.\\
substantial & Permanent unavailability of digital services or cooling results in a total economic loss, privacy issues.\\
threatening & High fines from violation of the law results in private insolvency, severe sickness from food poisoning.\\
		\hline		
		\end{tabular}
\end{table}

From Table~\ref{tab:harm} we can derive the protection needs of the data. In Table~\ref{tab:protection}, we have analyzed if an interruption, interception, modification or fabrication of data has an impact on integrity, availability or confidentiality. If this impact can produce negligible or limited harm, the protection need is normal. If it can be substantial, the protection need is high. If the harm can be threatening, the protection need is very high. 

\begin{table}[H]
		\caption{\label{tab:protection} \uppercase{Protection Needs}}
	\centering
		\begin{tabular}[h]{|c|c|c|c|}\hline
\textsl{Data} & \textsl{Integrity} & \textsl{Availability} & \textsl{Confidentiality} \\\hline
D1 & normal & normal & high\\
D2 & normal & normal & high\\
D3 & normal & normal & high\\
D4 & high & high & high\\
D5 & normal & normal & high\\
D6 & high & high & high\\
		\hline		
		\end{tabular}
\end{table}

D1--D5 might allow to infer personal information, e.g., eating habits, preferred foods, the daily routine or if the user is sick or on vacation. Thus, D1--D5 have "high" protection needs for confidentiality. 
The security of the user's network and the functionality of the fridge depend on D4 and D6. A misconfiguration, a disclosure of passwords, a manipulated OS update or an attacker knowing the patch-level of the software might result in a substantial harm (cf. Table~\ref{tab:harm}). Thus, D4 and D6 have the protection need "high" for all three dimensions. In our domestic setting, a threatening harm is highly improbable, and we do not assign the protection need "very high". This may be different in other scenarios, e.g., if a hospital uses the smart fridge to cool medications. 

The protection needs are inherited from the data to each IT asset managing the data, as listed in the asset matrix Table~\ref{tab:matrix}.
The maximum principle requires that an asset is assigned with the highest protection need of all data it uses. 
For example, the vendor's cloud (A3) is part of the use case "Update" (U8), which includes data of the operation system (D6) with the protection need "high" for the protection dimensions integrity, availability and confidentiality. 
Thus, even if A3 handles less sensible data (D1 and D2), the protection need of A3 is "high" for each protection dimension. 

From the asset matrix it follows that \textit{any} device, network connection and organization need to maintain a "high" level of protection for each dimension, because either D4 or D6 is part of any use case. 
As consequence from the asset matrix, it is problematic to use a smart fridge as a multimedia hub in a smart home as well. 
Because the fridge runs services with "high" protection needs, the much less sensitive media-playback service must be secured at level "high" as well.

% % % % % % % % % % % % % % % % % % % % % % % % % % % % % % % % % % % % % % % % % % % % % % % % % % % % % % % % % % % % % % % % % % % % % % % % 
\subsection{Long-Term Risks}

\begin{table*}[htb]
		\caption{\label{tab:compliance} \uppercase{Long-Term Compliance Risks}}
	\centering
		\begin{tabular}[htb]{|p{2cm}|p{1.0cm}|p{1.4cm}|p{1.3cm}|p{9.0cm}|}\hline
\textsl{Risk} & \textsl{Orga.} & \textsl{Devices} & \textsl{Connections} & \textsl{Description}\\\hline
Privacy & O2-O5 & A2-A5 & C2-C5 & Changing legislation, new codes of conduct, etc. impose limitations on the exchange of personal data with certain countries or parties \cite{mccullagh2017brexit}. \\
Global Rules & O2-O5 & A2-A4 & C2-C4 & New trade restrictions, sanctions, technology bans etc. restrict the use of an asset from certain countries or parties \cite{ziye2020china}.\\
Local Rules & O1-O5 & A1-A5 & C1-C5 & Local regulations, e.g., from environmental protection, consumer protection  or electromagnetic compatibility, restrict the use of an asset \cite{ecodesign}.\\
Expiration & O2-O5 & A1-A5 & C1-C5 & Disagreements to common compliance standards, expired certifications or approvals, non-renewed audits, etc., render the involved asset untrusted \cite{mak1996differences}. \\
Concealment & O1-O5 & A1-A5 & C1-C5 & Unknown characteristics at time of purchase disallow the further use of an asset, e.g., if it turns out that a build-in camera falls under espionage acts \cite{cayla2022}.\\
		\hline
		\end{tabular}
\end{table*}
\begin{table*}[htb]
		\caption{\label{tab:economic} \uppercase{Long-Term Economic Risks}}
	\centering
		\begin{tabular}[htb]{|p{2cm}|p{1.0cm}|p{1.4cm}|p{1.3cm}|p{9.0cm}|}\hline
\textsl{Risk} & \textsl{Orga.} & \textsl{Devices} & \textsl{Connections} & \textsl{Description}\\\hline
Degradation & O2-O5 & A3, A4 & C2-C4 & The service quality of an asset might be reduced, e.g., to nudge customers into new services by delaying updates or reducing performance of old services \cite{lyons2012net}.\\
Licensing & O2-O5 & A1, A3, A4 & C2-C4 & The revenue model might change. For example, an organization might switch its services to a pay-per-use model for an asset \cite{cusumano2008changing}.\\
Discontinuation & O2-O5 & A3, A4 & C2-C5 & One of the parties involved discontinues its service or makes it unattractive to use it from an economic point of view \cite{lemley2018essential}.\\
Liabilities & O2-O5 & A3, A4 & C2-C5 & One of the parties involved discontinues its business, and its contractual liabilities become void at once \cite{schwartz1985products}.\\
		\hline		
		\end{tabular}
\end{table*}
\begin{table*}[htb]
		\caption{\label{tab:operational} \uppercase{Long-Term Operational Risks}}
	\centering
		\begin{tabular}[htb]{|p{2cm}|p{1.0cm}|p{1.4cm}|p{1.3cm}|p{9.0cm}|}\hline
\textsl{Risk} & \textsl{Orga.} & \textsl{Devices} & \textsl{Connections} & \textsl{Description}\\\hline
Inflexibility & O1-O5 & A1-A5 & C1-C5 & Due to missing functional updates, it becomes challenging to connect an asset to recent services or devices \cite{mutchler2016target}.\\
Unreliability & O2-O5 & A1-A5 & C1-C5 & The service level in terms of reliability, throughput, etc. of the asset degrades, e.g., due to reduced support for legacy products \cite{ford2012icebergs}.\\
Unmaintainability & O2 & A1 & C1, C3, C5 & Due to the use of outdated interfaces and closed-source components it becomes difficult to find manuals, experts or spare parts to that maintain the asset \cite{ferreira2017discontinued}.\\
Insecurity & O2 & A1 & C1, C3, C5 & Without security updates and by using out-of-date security protocols, the asset cannot be operated any more \cite{ford2012icebergs}.\\
Defectiveness & O2 & A1 & C1, C3, C5 & Modernizations in the IT ecosystem make technical debts of an asset visible, e.g., if a network protocol uses bits that were reserved for future use \cite{kruchten2012technical}. \\		\hline		
		\end{tabular}
\end{table*}

After having identified the potential harms and the assets in our IT ecosystem that need special protection, we can compile IT-security risks. 
In order to obtain a comprehensive set of risks, we inspect each asset (organization, device, connection) in isolation, and we look for reasons why, at some point in the future, the asset in question will no longer operate as it did at the time of purchase. Recall that we are specifically interested in long-term risks of the smart fridge. Thus, in line with Step~3 of our research method, we filter out any potential risk that (a) is apparent at the time of purchase, or (b) is identical for a traditional fridge and a smart fridge. For example, we do not consider risks, such as the smart fridge is delivered with a pre-installed virus in its operating system, or the cooling unit fails after some time. 

The resulting set of risks is long and repetitive, because some risks materialize across different assets. For example, licensing risks due to changing revenue models can affect many assets and network connections at some time in the future, and occur at multiple organizations. For this reason, we need Step~4 of our research method to consolidate risks. 

Tables~\ref{tab:compliance}-\ref{tab:operational} show our consolidated set of long-term risks for 8 use cases for smart fridges. 
To our surprise, many of those risks are identical to the risks, which we had exemplarily identified for a single artefact (the network connection between a smart device and a cloud server) in~\cite{buchmann2020identifying}. This confirms the reproducibility of our research method. 

We have structured our set of risks into three groups:
\textbf{Long-term compliance risks} are produced by changing local, national or international rules and standards. Risks from this group mean that using the smart fridge may violate regulatory requirements in the future, even if it has fully complied with them at the time of purchase.
\textbf{Long-term economic risks} are the result of future business decisions of the organizations involved. Seven of the identified use cases require a complex IT ecosystem, as shown in our asset matrix (Table~\ref{tab:matrix}). If an organization ceases operation or moves to a different revenue model in the future, the remaining IT ecosystem may no longer able to support all use cases in an economic manner. This also includes a pay-per-security-update model.
Finally, \textbf{long-term operational risks} consider the technical perspective of operating a smart fridge together with its IT ecosystem for more than 10 years. 
Operational risks include technical challenges when trying to connect an outdated device to a new one, and maintenance issues due to missing experts and spare parts for the IT ecosystem needed. 

The tables only specify risks that impede a smart fridge, even if the same risk may be also associated with other devices in our IT ecosystem.
For example, risk "Unmaintainability" is listed for device~A1 (the fridge itself) and organization~O2 (the fridge's vendor), although the same risk exists for any other IT asset that is used for a decade or more.

Note that even "cooling" (U1), the most basic use case, is at risk in the long run. A smart fridge may have a reduced control panel. Such fridges depend on the use case "remote" (U6), which needs an Internet connection to the vendor's cloud and a smartphone app. For example, the Samsung RF27T5501SG requires the user to download the "SmartThings" app and register for the Samsung cloud with a personalized account.

\section{Conclusion}
\label{sec:conclusion}

When non-smart devices are replaced by smart ones, the integrated IT components generate new risks, that may limit the operational life-span of the smart device unexpectedly.
Such risks originate from different life cycles of digital and physical objects, from changing legislation, from future business decisions by the parties involved and from the technical complexity of the IT ecosystem needed. 

In this paper, we have compiled a catalog of long-term risks for smart fridges. Our catalog consists of risks, which might materialize years after the purchase. The risks are specific to the smart device, i.e., we have omitted any traditional risk that also exists for conventional fridges. 
Because we have used a well-structured research method, we think that our risk catalog is exhaustive for compliance risks, economic risks and operational risks. 
Our risk catalog can be adapted to many use cases and smart devices that use a similar IT architecture.

\section*{Acknowledgment}
% \addcontentsline{toc}{section}{Acknowledgment}

We would like to thank Badr Aldin Saada for his outstanding help and support with our risk analysis.

%\newpage
%\listofnotes

%\printbibliography

\bibliographystyle{IEEEtran}
\bibliography{paper}

\end{document}